\documentclass[twocolumn,aps]{revtex4}

\usepackage{graphicx}

\begin{document}

\title{
Coherent spin transport through a 350-micron-thick Silicon wafer
}

\author{Biqin Huang}
\altaffiliation{bqhuang@udel.edu} \affiliation{Electrical and
Computer Engineering Department, University of Delaware, Newark,
Delaware, 19716}
\author{Douwe J. Monsma}
\affiliation{Cambridge NanoTech Inc., Cambridge MA 02139}
\author{Ian Appelbaum}
\affiliation{ Electrical and Computer Engineering Department,
University of Delaware, Newark, Delaware, 19716}

\begin{abstract}
We use all-electrical methods to inject, transport, and detect spin-polarized electrons vertically through a 350-micron-thick undoped single-crystal silicon wafer. Spin precession measurements in a perpendicular magnetic field at different accelerating electric fields reveal high spin coherence with at least 13$\pi$ precession angles. The magnetic-field spacing of precession extrema are used to determine the injector-to-detector electron transit time. These transit time values are associated with output magnetocurrent changes (from in-plane spin-valve measurements), which are proportional to final spin polarization. Fitting the results to a simple exponential spin-decay model yields a conduction electron spin lifetime ($T_1$) lower bound in silicon of over 500ns at 60K. 
\end{abstract}

\maketitle

Silicon (Si) has been broadly viewed as the ideal material for spintronics due to its low atomic weight, lattice inversion symmetry, and low isotopic abundance of species having nuclear spin.\cite{ZUTICRMP, ZUTICPRL, LYON} These qualities are in contrast to the high atomic weight, inversion-asymmetric zinceblende lattice, and high nuclear spin of the well-studied semiconductor GaAs,\cite{KIKKAWA1, KIKKAWA2, CROWELL1, CROWELL2, JIANG1, JIANG2, BHATTACHARYA} which consequently has a relatively large spin-orbit and hyperfine interaction.\cite{ZUTICRMP} The resulting long spin lifetime and spin coherence lengths in Si may therefore enable spin-based Si integrated circuits.\cite{ZUTICNV, QC} 

Despite this appeal, however, the experimental difficulties of achieving coherent spin transport in silicon were first overcome only recently, by using unique spin-polarized hot-electron injection and detection techniques.\cite{APPELBAUMNATURE, BIQINJAP, 35percentAPL, SPINFETEXPT} (Subsequently, tunnel spin injection was demonstrated using optical detection with circular polarization analysis of weak indirect-bandgap electroluminescence.\cite{JONKERNATPHYS}) In Refs. \cite{APPELBAUMNATURE} and \cite{BIQINJAP}, spin transport through 10 $\mu$m of silicon was demonstrated and a spin lifetime lower bound of $\approx$1 ns at 85K was estimated. Using a new type of hot-electron spin injector that gives higher spin polarization and output current, we now show that (like in GaAs)\cite{KIKKAWA2} coherent spin transport can be observed over much longer lengthscales: we demonstrate transport vertically through a 350 $\mu$m-thick silicon wafer, and derive a spin lifetime of at least 500 ns at 60K (two orders of magnitude higher than metals or other semiconductors such as GaAs at similar temperature\cite{KIKKAWA1, AWSCHALOMLOSSSAMARTH}).
 
\begin{figure}
  \includegraphics[width=7.5cm,height=7cm]{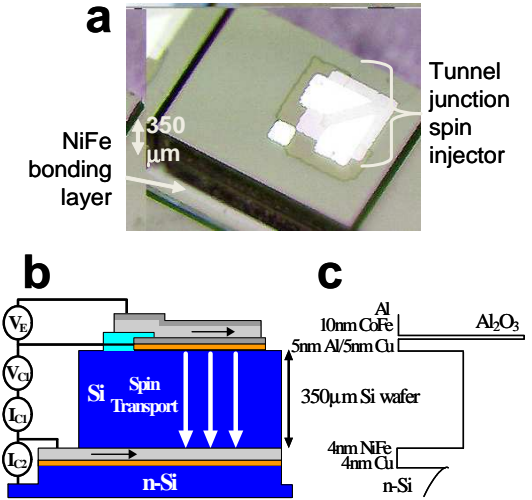}
  \caption{(a) Photograph of one of our 350-micron Si spin transport devices in perspective, showing the hot-electron spin-injection structure on top. The lateral size of the device is approximately 1.4 mm$\times$1 mm. (b) Schematic side-view with electrical configuration shown, and (c) associated conduction band diagram and constituent layers. Spin-polarized electrons are injected from the top of the device and drift in an electric field to the bottom where they are detected with a semiconductor-ferromagnetic metal-semiconductor structure employing spin-dependent inelastic scattering.}
\end{figure}

As in Refs. \cite{APPELBAUMNATURE, BIQINJAP, 35percentAPL, SPINFETEXPT}, we use ultra-high vacuum metal-film wafer bonding\cite{SVTSCIENCE} to build a semiconductor-metal-semiconductor hot-electron spin detection structure. A 2" diameter double-side polished 350-micron-thick undoped (resistivity $>$ 13 k$\Omega\cdot$cm @ room-temperature) single-crystal Si(100) wafer is bonded to a 1-10  $\Omega\cdot$cm n-type Si(100) wafer with a 4nm Ni$_{80}$Fe$_{20}$/ 4nm Cu bilayer. \cite{SVTSCIENCE, JANSEN} This process began with wafer cleaning in buffered HF solution and immediate loading into our wafer-bonding chamber. After pump-down and bakeout to the base pressure of 1E-8 Torr, 4nm of Cu was thermally evaporated onto the n-Si wafer only. (This layer is necessary to reduce the hot-electron collector Schottky barrier height there.)\cite{SVTPRL} During subsequent thermal evaporation of 2nm Ni$_{80}$Fe$_{20}$ on both wafers, the ultra-clean surfaces of the deposited metal films were pressed together {\it in-situ} with nominal force, forming a cohesive bond with a re-crystallized 4nm-thick Ni$_{80}$Fe$_{20}$ layer. \cite{SVTSCIENCE} 

Although these bonding steps are identical to our previous reports with 10 $\mu m$-thick transport layer devices, the subsequent procedure used to fabricate 350 $\mu m$-thick transport layer devices differs significantly. In the present work, the outside polished surface of the undoped Si wafer in the bonded pair was covered by a protective 1 $\mu$m-thick SiO$_2$ layer deposited by an electron-beam source. A wafer saw was used to first cut through the undoped Si wafer and buried metal bonding layer, partially through the n-Si wafer to define individual device mesas. Then, the saw was used to cut trenches in the undoped Si wafer close to, but not through, the buried metal bilayer. Wet chemical etching with tetramethyl ammonium hydroxide (TMAH) removed the remaining Si and exposed the buried Ni$_{80}$Fe$_{20}$ for electrical contact.\cite{SVPT} After protective SiO$_2$ removal with buffered HF, a 40nm Al/10nm Co$_{84}$Fe$_{16}$/Al$_2$O$_3$/5nm Al/5nm Cu tunnel junction hot-electron spin injector was deposited using electron-beam evaporation through shadow masks for lateral patterning.\cite{35percentAPL} 

Figure 1(a-c) illustrates the geometry of one of our completed four-terminal silicon spin-transport devices. The optical image in Fig. 1(a) shows a device (before contacting with wire-bonds) having a lateral size of approximately 1$\times$1.4mm. The schematic side-view and associated conduction band diagram in Fig. 1(b) and (c), respectively, shows the vertical geometry and can be used to elucidate the means of spin injection and detection. When a voltage bias $V_E$ is applied across the emitter tunnel junction, electrons that are spin polarized at the cathode Co$_{84}$Fe$_{16}$/Al$_2$O$_3$ interface tunnel through the oxide barrier and some travel ballistically through the nonmagnetic Al/Cu anode bilayer. Those electrons with energy above the Cu/Si Schottky barrier ($\approx$0.6eV)\cite{SZEBOOK} can couple with Si conduction band states and then quickly thermalize to the conduction band minimum.\cite{SIBEEM} These spin-polarized electrons are then accelerated in an applied electric field vertically through the 350 micron-thick wafer and toward the opposite side of the undoped Si, where they are ejected from the conduction band into the buried metal layer. Because the ferromagnetic Ni$_{80}$Fe$_{20}$ layer has a spin-dependent bandstructure, the inelastic scattering rates of these hot electrons to the Fermi energy is also spin-dependent. Therefore, the number of ballistic electrons that can couple with conduction band states in the n-Si collector on the other side (forming the ``second collector current'' $I_{C2}$) is dependent on the relative orientation of final spin direction and ferromagnet (FM) magnetization. 

The spin-polarized electron injector we use here is notably different from the design in previous studies, where spin-dependent scattering in the base anode (ballistic spin filtering) was the operating mechanism.\cite{APPELBAUMNATURE, BIQINJAP, SPINFETEXPT} In the devices used in the present work, initial spin polarization is obtained by direct tunneling from the cathode FM (Co$_{84}$Fe$_{16}$) through the Al$_2$O$_3$ tunnel junction oxide. This design gives several advantages: 1. the FM is removed from the Si surface, preventing the formation of a non-magnetic silicide having strong, randomly-oriented magnetic moments. The elimination of this ``magnetically-dead'' region (which could cause significant spin scattering) maintains a high initial spin polarization.;\cite{35percentAPL} 2. Ballistic hot-electron transport before injection into the Si conduction band is through non-magnetic Al and Cu, which have much larger ballistic mean-free-paths than typical FMs, resulting in higher injected current ($I_{C1}$) and the spin-signal output current ($I_{C2}$) it drives; and 3. The Cu/Si Schottky barrier height is relatively low,\cite{SZEBOOK} further increasing $I_{C1}$.

\begin{figure}
  \includegraphics[width=6.5cm,height=5.75cm]{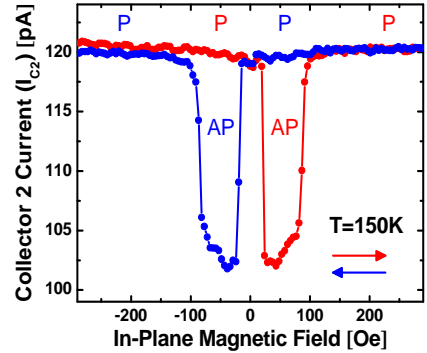}
  \caption{In-plane magnetic hysteresis measurement of second collector current $I_{C2}$ at constant emitter voltage bias $V_E=-1.3$V and constant accelerating voltage $V_{C1}=20$V, showing $\approx$18\% spin-valve effect at 150K. The labels ``P'' and ``AP'' refer to parallel and antiparallel injector/detector magnetization configuration, respectively. Injected current $I_{C1}$ is 6.6$\mu$A.}
\end{figure}

If spin ``up'' is both injected and detected with parallel FM magnetizations (and no spin flipping or rotating process occurs in the Si bulk) a relatively high $I_{C2}$ should be measured. On the other hand, if spin ``up'' is injected, but spin ``down'' is detected (with anti-parallel FM magnetizations), $I_{C2}$ will be relatively lower, again assuming no spin flips or rotations. The ferromagnetic layers chosen for the injector (Co$_{84}$Fe$_{16}$) and detector (Ni$_{80}$Fe$_{20}$) have different coercive (or switching) fields, which enables external control over the relative orientation of spin injection and detection axes with an in-plane magnetic field. At 150K, clean spin-valve signals at constant emitter bias $V_E=-1.3$V and accelerating voltage $V_{C1}=20$V (resulting in $\approx$580 V/cm electric field)\cite{BIQINJAP} indicate a $\approx$18\% change in $I_{C2}$ when the magnetizations of injector and detector are switched from a parallel (P) to anti-parallel (AP) configuration by an externally-applied in-plane magnetic field, according to our expectations (as shown in Fig. 2). This magnetocurrent ratio ($MC=(I_{C2}^P-I_{C2}^{AP})/I_{C2}^{AP}$) corresponds to an electron current spin polarization of approximately $\mathcal{P}=MC/(MC+2)\approx$ 8\%.\cite{SPINFETEXPT} However, this evidence for spin transport is not conclusive without observation of spin precession and dephasing (Hanle effect\cite{JOHNSON1985, JOHNSON1988}) in a perpendicular magnetic field. \cite{MONZON}

\begin{figure}
  \includegraphics[width=6.5cm,height=10cm]{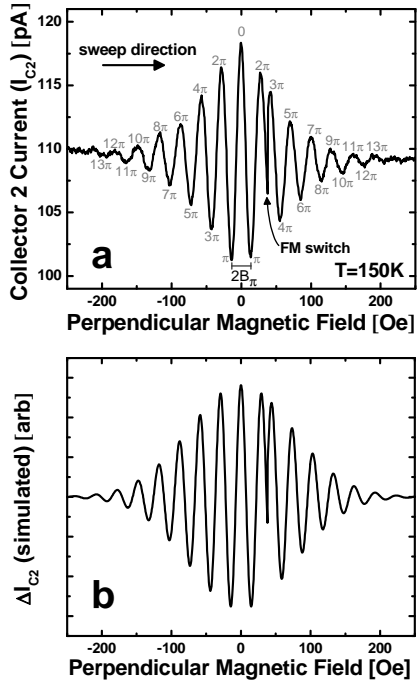}
  \caption{(a) Spin precession and dephasing (Hanle effect) of Si conduction-band electrons in a perpendicular magnetic field at 150K using the same voltage bias conditions as in Fig. 2, showing up to 13$\pi$ rad precession angles. The ``FM switch'' is caused by a residual in-plane magnetic field component switching the in-plane magnetization of the Ni$_{80}$Fe$_{20}$ detector at $\approx$+38 Oe, which inverts the maxima and minima at higher positive field values. (b) Simulation of the measurement in (a), using the drift-diffusion spin precession model given by Eq. \ref{IC2EQN}.}
\end{figure}

A perpendicular magnetic field $\vec{B}$ exerts a torque $(g \mu_B/\hbar)\vec{S} \times \vec{B}$ on the electron spin $\vec{S}$, causing spin rotation (precession) about $\vec{B}$. Here, $g$ is the electron spin g-factor, $\mu_B$ is the Bohr magneton, and $\hbar$ is the reduced Planck constant. Our spin detector measures the projection of final spin angle on an axis determined by the Ni$_{80}$Fe$_{20}$ magnetization, so we observe oscillations in $I_{C2}$ as the precession frequency $\omega=g\mu_B B/\hbar$ is varied.  

Fig. 3(a) shows our measurement of $I_{C2}$ in varying perpendicular magnetic field with the same temperature and bias conditions as in Fig. 2. The measurement begins at negative field values when the injector/detector magnetizations are in a parallel orientation. As the field is increased, we see multiple oscillations due to spin precession. However, when the field reaches $\approx$+38 Oe, a small in-plane component of the applied field switches the magnetization of the magnetically softer Ni$_{80}$Fe$_{20}$, resulting in an antiparallel injector/detector orientation that inverts the magnitudes of maxima and minima.

The final precession angle $\theta$ at the detector is simply the product of transit time from injector to detector, $\tau$, and spin precession frequency $\omega$. Since our measurement is an average of the precession angles over all electrons arriving at the detector regardless of transit time $\tau$, the magnitudes of higher-order extrema (labeled in Fig 3(a)) are reduced by the dephasing associated with a distribution in transit times $\Delta \tau$ caused by random diffusion. 

We can simulate our measurement in the presence of both drift and diffusion by integrating the contributions to our signal from an ensemble of precessing spins with a diffusion-controlled distribution of transit times using a simple model\cite{CROWELL2, SPINFETTHEORY}:

\begin{equation} 
\label{IC2EQN}
\Delta I_{C2} \sim \int_0^{\infty}\frac{1}{2\sqrt{\pi D t}}e^{-\frac{(x-vt)^2}{4Dt}}\cdot \cos(\omega t)\cdot e^{-t/\tau_{sf}}dt,
\end{equation} 

\noindent where $D$ is the diffusion constant, $v$ is drift velocity, and $\tau_{sf}$ is effective spin lifetime. The integrand is simply the product of the effects of drift and diffusion, precession, and finite spin lifetime. Using $x=L=350\mu$m, $D=200$ cm$^2$/s, $v=2.9\times 10^6$ cm/s,\cite{CANALI} and $\tau_{sf}=73$ ns (see below), we find excellent agreement between experiment and model in Fig. 3(b). (In this simulation, the sign is inverted for magnetic field values $>$38 Oe to match the experimental results.) 

Despite transport through 350 microns of undoped Si, high spin coherence with at least $13\pi$ spin precession angle (more than six full rotations) is evident in Fig. 3(a), which is even greater than what was previously demonstrated using a much shorter 10 $\mu m$-thick transport layer.\cite{APPELBAUMNATURE} Because the transit time is therefore much longer in the thicker devices, it could be argued that diffusion should play a larger role and dephasing should suppress multiple oscillations in precession measurements. The results of the experiment and consistent model simulation clearly conflict with this reasoning. 

The somewhat counterintuitive result can be explained with a simple argument: If transport is dominated by drift in the applied electric field\cite{BIQINJAP}, the transit time is given by $\tau=L/v=L^2/(\mu V_{C1})$, where $\mu$ is the electron mobility, $L$ is the transport length, and $v$ is drift velocity.\cite{SPINFETTHEORY} The width $d$ of an initially injected infinitesimally-narrow gaussian spin distribution will increase by diffusion during this transit time to $d=\sqrt{D\tau}=L\sqrt{D/(\mu V_{C1})}$. Since the width of the distribution of transit times $\Delta \tau$ is $d/v$, the relative uncertainty in the distribution of final precession angle $\theta$ at the detector is $\Delta\theta/\theta=\frac{\omega \cdot \Delta \tau}{\omega\tau}= \sqrt{D/(\mu V_{C1})}$. This result is independent of the transit length $L$, so we can expect the same amount of dephasing regardless of the distance from injector to detector for any fixed precession angle (assuming ohmic behavior, $v=\mu E$, where $E$ is internal electric field).  

From the oscillation period of spin precession measurements ($2B_\pi $, as shown in Fig 3(a)), we can determine the average spin transit time in any given accelerating electric drift field conditions (induced by $V_{C1}$) through $\tau=h/(2g\mu_B B_\pi)$. The normalized magnetocurrent $\Delta I_{C2}/I_{C1}$ determined by spin-valve measurements like those in Fig. 2 gives a quantity that is proportional to conduction electron current spin polarization, $\mathcal{P}$.\cite{APPELBAUMNATURE,BIQINJAP} Associating this value with the transit times given by precession measurements (see above) gives data which can be fit with a simple exponential decay model, where

\begin{equation}
\label{EXPEQN}
\mathcal{P} \propto exp(-\tau/T_1).
\end{equation}

\noindent The timescale $T_1$ is the longitudinal spin lifetime, since our spin-polarization data is derived from spin-valve measurements with in-plane magnetic fields colinear to the spin direction.

\begin{figure}
  \includegraphics[width=6.25cm,height=10.5cm]{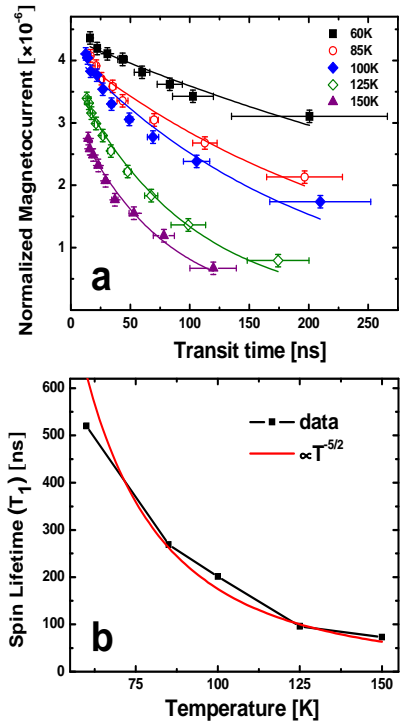}
  \caption{(a) Fitting the normalized magnetocurrent ($\Delta I_{C2}/I_{C1}$) to an exponential decay model (Eq. \ref{EXPEQN}) using transit times derived from spin precession measurements (like those in Fig. 3(a)) at variable internal electric field yields measurement of longitudinal spin lifetimes ($T_1$) in undoped bulk Si. (b) The experimental $T_1$ values obtained as a function of temperature are compared to Yafet's $T^{-5/2}$ power law for indirect-bandgap semiconductors\cite{YAFET}.}
\end{figure}

The best-fits to our data at 60K, 85K, 100K, 125K and 150K using the expression in Eq. \ref{EXPEQN} are 520ns, 269ns, 201ns, 96ns, and 73ns, respectively, as shown in Fig. 4(a) and (b). These lifetimes are much greater than the $\approx$1 ns lifetime lower bound estimated in previous work, because with the much longer transit lengths here, the applied accelerating voltage $V_{C1}$ varies the transit time over a range of $\approx$200 ns; previously the range was only several hundred ps, and parasitic electronic effects suppressed our estimate.\cite{APPELBAUMNATURE, BIQINJAP} The temperature dependence of these spin lifetimes fit well to the expected behavior in an indirect-bandgap semiconductor predicted by Yafet ($\propto T^{-5/2}$), as shown in Fig. 4(b).\cite{YAFET, LEPINE, JAROSLOVACA} The relative absence of other relaxation mechanisms in Si is responsible for the long spin lifetimes.

Certainly, higher temperature operation is desirable. However, thermionic leakage at the second collector Schottky barrier and the difficulties of reliably operating our tunnel junction spin injector at high voltages necessary are the present limitation to increasing this temperature. Although observation of spin precession at high electric fields are possible at lower temperatures, measurements of spin lifetime below 60K are currently prevented by carrier freeze-out effects.

The long lifetimes measured here are lower bounds, with the possibility that parasitic electronic effects artificially suppress the values obtained.\cite{BIQINJAP} Hence, spin lifetimes could be higher with associated longer transport lengths. Due to the thickness limitations of Si wafers, we will explore these longer distances with lateral transport devices. This achievement should enable true spintronic circuits intimately compatible with existing Si CMOS logic, and potentially extend the performance trend of Si devices beyond its limits set by conventional approaches.

We gratefully acknowledge J. Fabian for helpful discussions. This work was supported by DARPA/MTO.

\end{document}